# Weather forecasting using Convex hull & K-Means Techniques – An Approach

Ratul Dey[a], Sanjay Chakraborty[b], Lopamudra Dey[c]

[a] Computer Science & Engineering, Institute of Engineering and Management, Kolkata, India, email: ratul170292@gmail.com
[b] Computer Science & Engineering, Institute of Engineering and Management, Kolkata, India, email: sanjay_ciem@yahoo.com
[c] Computer Science & Engineering, Heritage Institute of Technology, Kolkata, India, email: lopamudra.dey1@gmail.com

*Abstract*

Data mining is a popular concept of mined necessary data from a large set of data. Data mining using clustering is a powerful way to analyze data and gives prediction. In this paper non structural time series data is used to forecast daily average temperature, humidity and overall weather conditions of Kolkata city. The air pollution data have been taken from West Bengal Pollution Control Board to build the original dataset on which the prediction approach of this paper is studied and applied. This paper describes a new technique to predict the weather conditions using convex hull which gives structural data and then apply incremental K-means to define the appropriate clusters. It splits the total database into four separate databases with respect to different weather conditions. In the final step, the result will be calculated on the basis of priority based protocol which is defined based on some mathematical deduction.

1. **Introduction**

In our human society, the crop is one of the essential food, production of the crop is directly depends on the weather. If we predict weather, it will be benefited our farmers and also makes an effect on the national growth. It's very important because it lets people know what is coming and how good/bad the weather will be. With the help of previous storage data, it is possible to predict weather. Analysis and probability give accurate result for prediction. Weather warnings are an important forecast as because it is used to protect human life and wealth. Today's advanced technology is used to find out the pattern of data which have an ability to predict the future atmosphere. Weather forecasting is directly dependent on the natural molecules of air like Nitrogen dioxide($NO_2$), Ozone($O_3$), Carbon Dioxide ($CO_2$), Sulfur dioxide ($SO_2$) etc. This paper focuses on specific zone 'Kolkata'. Use of some specific models or techniques is to find out the tendency of the weather on that particular zone. In a few prediction operations, some advanced numerical analysis has been used.

This paper has been organized in the following way, section 2 deals with some previous work used to predict weather conditions. Section 3deals with those basic techniques which are applied in this paper helps to properly predict the weather conditions of 'Kolkata' city. Then section 4 defines the different matters and their effects on our weather. Section 5 and 6 represents the actual proposed technique and its algorithm elaborately. Then section 7 shows the application of this new approach on a set of air pollution data and analyse the resultant prediction. At last section 8 describes the conclusion and future scope of this approach.

2. **Related Work**

There are several applications that are used for weather prediction. The same approach is also used in population forecasting, voting forecasting, etc. use K-Means clustering of air pollution database and the list of weather category will be developed based on maximum mean values of the cluster. If some new data come, then use incremental K-means of those data [1]. The use of incremental K-means clustering algorithm, it evaluates the particular point of change in the database which performs better than the existing K-means clustering by changing threshold values [2]. Incremental K-means clustering algorithm which is applicable in the periodically incremental environment and dealing with a bulk number of updates [1][2]. K-Means clustering algorithm is applied to a dynamic database where the data may be frequently updated [3]. Weather forecasting also be done by using artificial neural network [8]. Some paper describes two R packages for probabilistic weather forecasting. It offers an ensemble of post processing via 'Bayesian Model Averaging', which implements the 'Geostatistical Output perturbation method' (GOP). Both packages include function for evaluating predictive performance in addition to model fitting and forecasting. [5] Some work is done in respect of time series forecasting through clustering. Some useful pattern in the form of curves facilitates the forecasting through linear regression by matching to the closest pattern to each time series that has to be predicted. Another paper used numerical probability 'linear regression method' by regression analysis. Factor analysis using the reductive method builds relationship within variables [9] [6]. Analysis the rainfall data, give pattern of the weather changing. Forecasts based on our seasonal + trend + cycle model in the time series forecast [13] statistical model for ice formation at a particular time is fixed for the stored data[14]. A fast and stable incremental clustering algorithm use computationally modest and imposes minimal memory requirements [4] The root mean square errors of the superensemble precipitation indicating a skilled forecast. Multimodel superensemble partitions the forecast time first 'training phase' second 'forecast phase'. This utilizes a least squares minimization of the difference between anomalies of the model[16].

## 3. Related Methods

### 3.1 Data mining

Data mining is a process which gives knowledge about the database. It applies more computational technique for statistics, machine learning and pattern reorganization of a large database.

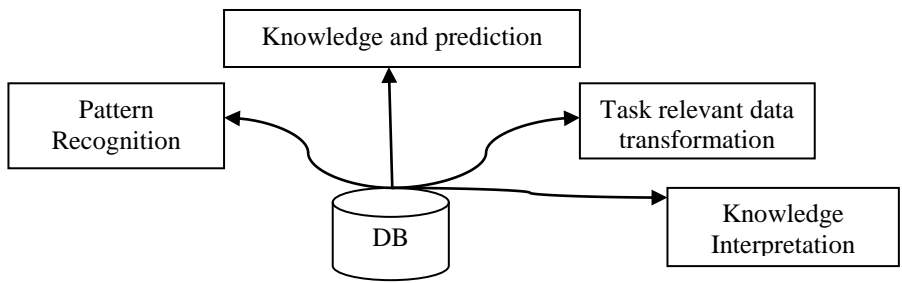

Figure I. Flow of data mining

### 3.2 Convex hull

This procedure ensures to take all different extreme values in the database. This is basically graph based calculation to find out the basic structure of the database. It is the smallest convex set of some given points. Firstly the point of the plot (P0, P1,……., Pn) on x,y plane graph. Choose left and right extreme point (PL, PR). P0is the starting point and it goes towards the immediate next point. If the next point goes towards the same direction or goes to right angle [Figure II] on the basis of previous direction then it will be accepted. If next point goes towards the left angle [Figure III] on the basis of previous direction, it rejects this node and previous node, and chooses different nodes, complete upper part of the graph. The same case is made in the lower part Figure IV. In large data analysis, it gives the fast average idea including all extreme data points.

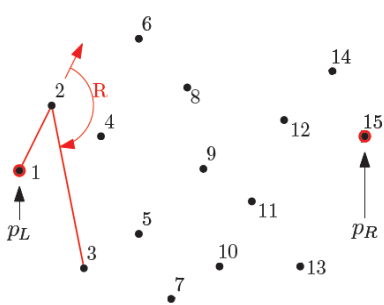 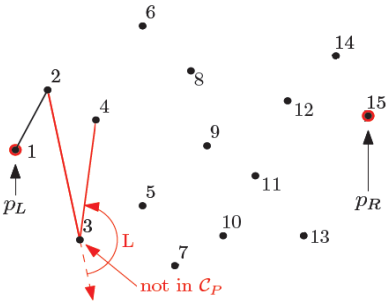 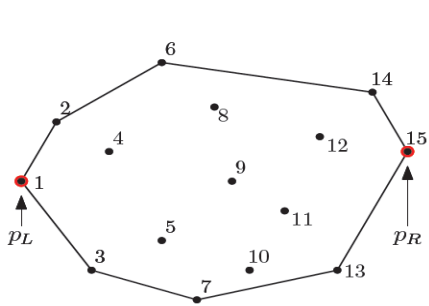

Figure II          Figure III          Figure IV

### 3.3. Incremental Clustering

Clustering is an unsupervised learning which represents the partitions of data which create groups of similar data. Its purpose is to find out underlying groups, or clusters which ideally share similar features. The purpose of unsupervised clustering is to learn how to use best discrete a value with the intent of classifying unseen data samples into several clusters with the assumption of features. Now some new data input to the database which can be inserted to the nearest cluster.

*3.4 Incremental* K-means

K means is a technique where count all values in the cluster. Firstly, identify the K cluster center; depending on the k Cluster centers create a cluster. Now the mean value of the cluster is called K-means. In Incremental K-means, data can be inserted at any time. When new data inserted into database, choose nearest cluster and it inserts to that cluster. Find out the new mean value of that cluster. Incremental clustering is designed using the cluster's metadata captured from the K-Means results.

4. **Air Molecules effect on weather**

$SO_2$: - $SO_2$ is produced by industrial processes, burning of coal and petroleum comes out the sulphur compound in the air. If it increases, the weather could be smog, fog, chance of Acid rain, if cold air present with calm winds and high humidity it may occur dense fog or Toxic Smog.

$NO_2$: Due to $No_2$ colour reddish brown its emitted high temperature combustion. $NO_2$, reaction with $O_2$, hv, etc.
$No_2 + O_2 \rightarrow No + O_3$ or $No_2 + hv \rightarrow No + O$ increase of No in the air molecules, weather may be cold and dry.

$CO_2$: - *Carbon* Dioxide is the most common pollutant element in the air molecules. Increase of $co_2$ reflects the weather as smoggy humid and hot. $CO_2 + 2 H^+ + 2 e^- \rightarrow CO + H_2O$ can create carbon monoxide, which equally affect the weather.

$O_3$ :- Ozone is more readily formed on warm, sunny days when the air is stagnant.
**PM 10**:- Particles between 2.5 and 10 micrometers in diameter are referred as PM10 which may increase dust in air molecules.

5. **Proposed Model**

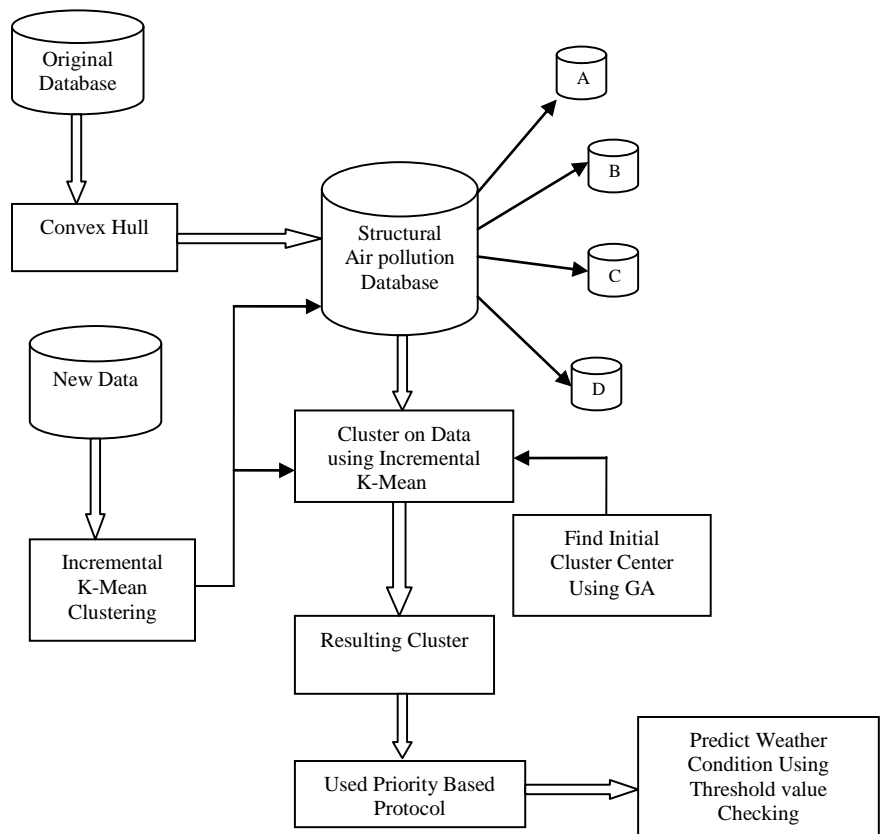

Figure V Proposed Model

Weather forecasting is expected to provide insights into future meteorological conditions for a specific locality, region and over a specified period of time. Basically the weather depends upon the air molecules which can absorb (high frequency sunray) temperature. [Figure V] At a regular interval of every one hour, the system collects air molecules data. The convex hull algorithm uses these raw data to yield information from the region value of large information to find 'Structural Air Pollution Database'. In the next step k-mean algorithm is applied to the above structural data base. After this, the resultant data are finally stored in the main database. Now the main database is divided into four regions depending on the direction of wind flow over the year. A (December, January, February), B (March, April), C (May, June, July), D (August, September, October, November). 'A' region is called winter region. 'B' and 'D' are called temperate region. 'C' is called summer region. When any data search is needed, it will be searched in its particular domain. The k-mean algorithm organizes data into clusters according to the region. For insertion of new data items, the incremental k-mean algorithm is used to enter data into the proper cluster. The initial cluster center is produced using the genetic algorithm. When the user feeds data to the system, the data is compared to the previous set of data using the priority based algorithm. The database stores multiple year data, now use the order of priority with the help of $[(1/3-\alpha), (\alpha), (1/3)]$. As per record of statistics, weather (Temperature and humidity) mainly depends on $2^{nd}$ last year. So choose priority $(1/3-\alpha)$ as the last year, $(\alpha)$ choose as $2^{nd}$ last year (highest priority) and $(1/3)$ choose as $3^{rd}$ last year. Prediction result calculate has been done on the basis of three years, where $\alpha$ is a constant variable.

6. **Proposed Algorithm**

*Step-1* Collect data ($NO_2$, $O_3$, $CO_2$, $SO_2$) in every one hour and store in the original database).

*Step-2* Using the Convex Hull Technique, after every two hours data received is converted into structural data. All extreme data are also included. Then store it in the modified 'structural air pollution database'.

*Step-3* 'Structural air pollution database' splits into four sub-databases on the basis of weather separation.

*Step-4* Apply K-Means clustering of the structural data, where initial cluster centers are guessed using well known genetic algorithm (GA).

*Step-5* If any new data insert into the database, then uses incremental K-Means clustering to accommodate with that new data insertion.

*Step-6* Finally, find the resulting clusters.

*Step-7* Result obtained in different years (max three years) can be predicted by priority based protocol.

*Step-8* Now to determine the probable weather Condition some well known threshold temperature value ranges can be used.

7. **Mathematical Illustration based on various Examples**

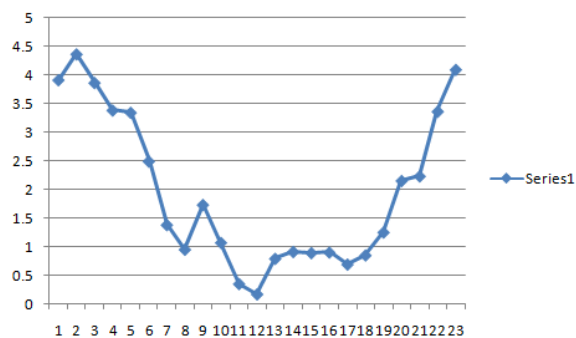

Figure VI (plot data on the graph hourly)

| Time | 01:00 | 02:00 | 03:00 | 04.00 | 05:00 | 06:00 | 07:00 | 08:00 | 09:00 | 10:00 | 11:00 | 12:00 |
|---|---|---|---|---|---|---|---|---|---|---|---|---|
| mg/m³ | 3.92 | 4.37 | 3.87 | 3.39 | 3.35 | 2.5 | 1.39 | 0.96 | 1.74 | 1.08 | 0.36 | 0.18 |

| Time | 13:00 | 14:00 | 15:00 | 16:00 | 17:00 | 18:00 | 19:00 | 20:00 | 21:00 | 22:00 | 23:00 |
|---|---|---|---|---|---|---|---|---|---|---|---|
| mg/m³ | 0.8 | 0.92 | 0.9 | 0.91 | 0.7 | 0.86 | 1.26 | 2.26 | 2.24 | 3.37 | 4.1 |

Table: 1 ('CO' value every one hour 02.01.2014)

Use Convex Hull to the plotted above graph, the value takes only outer region (3.92, 3.39, 1.39, 0.96, 0.36, 0.18, 0.7, 0.86, 1.26, 2.24, 4.1, 4.37 )

So the value of 'CO' on 02.01.2014 = (3.92 + 3.39 + 1.39 + 0.96 + 0.36 + 0.18 + 0.7 + 0.86 + 1.26 + 2.24 + 4.1 + 4.37 ) / 12
= 1.97

$$\text{Approx Calculated value} = \frac{Sum\ of\ outer\ region\ point}{Total\ number\ of\ outer\ region\ point}$$

Initially Table 1:- observe air molecules (carbon monoxide) value of every one hour, then plot into the graph, choose only outer region point using Convex Hull. Take the average value of the region and store in the next table (Table II).

| DATE | CO | $NO_2$ | $O_3$ | PM10 | $SO_2$ | Probable temp range °C |
|---|---|---|---|---|---|---|
| 01.01.14 | 1.87 | 162.85 | 17.67 | 234.666 | 6.40 | 24-27 |
| 02.01.14 | 1.97 | 186.97 | 16.75 | 226.01 | 5.839 | 24-27 |
| 03.01.14 | 1.359 | 114.589 | 18.28 | 168.88 | 4.95 | 22-26 |
| 04.01.14 | 0.925 | 89.62 | 28.43 | 150.09 | 4.45 | 22-26 |
| 05.01.14 | 1.50 | 151.34 | 21.131 | 213.363 | 4.64 | 22-26 |
| 06.01.14 | 1.58 | 148.31 | 10.54 | 212.01 | 4.717 | 22-26 |
| 07.01.14 | 1.555 | 121.19 | 12.199 | 199.87 | 5.23 | 22-25 |
| 08.01.14 | 0.92 | 90.51 | 24.15 | 137.675 | 4.13 | 22-26 |
| ………. | ………. | ………. | ……….. | ……….. | ………. | ………. |

Table: 2 (Daily data store)

After receiving all data, create clusters according to cluster center.

| Cluster Id | Cluster range of $CO_2$ | Cluster range of $NO_2$ | Cluster range of $O_3$ | Cluster range of PM10 | Cluster range of $SO_2$ |
|---|---|---|---|---|---|
| Cluster 1 | 0.5-1.75 | 50-105 | 5-12 | 50-121 | 2.55-3.39 |
| Cluster 2 | 1.75-2.5 | 105-177 | 12-19 | 121-168 | 3.39-4.5 |
| Cluster 3 | 2.5-3.55 | 177-190 | 19-23 | 168-204 | 4.5-5.6 |
| Cluster 4 | 3.55-4.10 | 190-210 | 23-28 | 204-229 | 5.6-6.8 |
| ……… | ……….. | …………. | ………… | ………… | ……….. |

Table: 3 Different ranges of air molecules based on various clusters

Four clusters are produced where each cluster depends on some range. Cluster shows the particular day is hot, fogy, hazy, dusty etc. Depend on the weather category and the basis of the regional database, it predicts temperature.

| Date | Probable temp range °C | Actual temp °C | Hit | miss | Weather category |
|---|---|---|---|---|---|
| 01.01.2014 | 24-27 | 26 | * |  | Smogy, fogs, haze and smoke |
| 02.01.2014 | 24-27 | 27 | * |  | Smog, fog, dusty and mist |
| 03.01.2014 | 22-26 | 23 | * |  | Dry, smog and mist |
| 04.01.2014 | 22-26 | 24 | * |  | Mist, haze and smoke |

Table: 4 Resultant Table (weather category, according to Cluster data)

**Accuracy Calculation**

$$\text{Accuracy} = \frac{Number\ of\ matched\ records}{Total\ number\ of\ records} \times 100$$

$$= \frac{284}{365} \times 100$$

$$= 78\% \text{ (approx)}$$

## 8. Conclusion and Future Work

In this paper, a new technique is introduced to predict the weather of upcoming days with the help of incremental K-means clustering algorithm and renowned convex hull technique. This approach is suitable for the dynamic databases where the climate data are changed frequently. It also uses the concept of genetic algorithm to make a probable guess of the initial cluster center to get more effective results. It includes rainfall analysis to give some better prediction. Here the concept is only introduced and mathematically analyzed with some basic example. As the future work, this approach will deal with some other air pollution databases of different regions to predict their weather and give a detail comparison among their weather conditions.

**Acknowledgements**

Special thanks to Prof. Dr. Partha Bhowmick from Indian Institute of Technology, Kharagpur whose comments improved the idea of this article.